\documentclass{elsart}

\usepackage{epsfig}
\newcommand{\be}{\begin{equation}}
\newcommand{\ee}{\end{equation}}
\newcommand{\bea}{\begin{eqnarray}}
\newcommand{\eea}{\end{eqnarray}}

\newcommand{\Fritiof}{\textsc{Fritiof}}

\newcommand{\UNIT}[1]{\mbox{$\,{\rm #1}$}}

\newcommand{\GeV}{\UNIT{GeV}}

\begin{document}

\begin{frontmatter}
\title{Low mass dilepton production at ultrarelativistic energies}
\author[FIAS]{E.L. Bratkovskaya,\corauthref{cor1}},
\ead{Elena.Bratkovskaya@th.physik.uni-frankfurt.de}
\corauth[cor1]{corresponding author}
\author[unig]{W.~Cassing}
\author[FIAS]{and O. Linnyk}
\address[FIAS]{Frankfurt Institute for Advanced Studies,
 Johann Wolfgang Goethe University,
  Ruth-Moufang-Str. 1,
 60438 Frankfurt am Main,
 Germany}
\address[unig]{Institut f\"ur Theoretische Physik, %
  Universit\"at Giessen,
  Heinrich-Buff-Ring 16,
  D-35392 Giessen, %
  Germany}

\begin{abstract}
Dilepton production in $pp$ and $Au+Au$ nucleus-nucleus collisions
at $\sqrt{s}$ = 200 GeV as well as in $In+In$ and $Pb+Au$ at 158 A$\cdot$GeV
is studied within the microscopic HSD transport approach.
A comparison to the data from the PHENIX Collaboration at RHIC shows
that standard in-medium effects of the $\rho, \omega$ vector
mesons  - compatible with the NA60 data for $In+In$ at 158
A$\cdot$GeV and the CERES data for $Pb+Au$ at 158 A$\cdot$GeV - do not
explain the large enhancement observed in the
invariant mass regime from 0.2 to 0.5 GeV in $Au+Au$ collisions
at $\sqrt{s}$ = 200 GeV relative to $pp$ collisions.
\end{abstract}


\end{frontmatter}

While the  properties of hadrons are rather well known in free space
(embedded in a nonperturbative QCD vacuum) the masses and lifetimes
of hadrons in a baryonic and/or mesonic environment are subject of
current research in order to achieve a better understanding of the
strong interaction and the nature of confinement. In this context
the modification of hadron properties in nuclear matter are of
fundamental interest (cf. Refs.
\cite{BrownRho,H&L92,Asakawa93,Shakin94,Klingl96}) since QCD sum
rules \cite{H&L92,Asakawa93,Leupold} as well as QCD inspired
effective Lagrangian models
\cite{BrownRho,Shakin94,Rapp,RappNPA,Peters} predict  significant
changes e.g. of the vector mesons ($\rho$, $\omega$ and $\phi$)
with the nuclear density $\rho_N$ and/or temperature $T$
\cite{CBRep98,rapp5}.

A  modification of vector mesons has been seen experimentally in
the enhanced production of lepton pairs above known sources in
nucleus-nucleus collisions at Super-Proton-Synchroton (SPS)
energies \cite{CERES,HELIOS}. As proposed by Li, Ko, and Brown
\cite{Li} and Ko et al. \cite{Li96} the observed enhancement in
the invariant mass range $0.3 \leq M \leq 0.7$ GeV might be due to
a shift of the $\rho$-meson mass following Brown/Rho scaling
\cite{BrownRho} or the Hatsuda and Lee sum rule
prediction~\cite{H&L92}.  The microscopic transport studies in
Refs. \cite{CBRep98,Cass95C,Brat97,Ernst} for these systems have
given support for this interpretation. On the other hand also more
conventional approaches that describe a melting of the
$\rho$-meson in the medium due to the strong hadronic coupling
(along the lines of Refs.~\cite{RappNPA,Peters}) have been found
to be compatible with the early CERES data
\cite{rapp5,Cass95C,CBRW97}. This ambiguous situation has been
clarified to some extent in 2006 by the NA60 Collaboration since
the invariant mass spectra for $\mu^+\mu^-$ pairs from In+In
collisions at 158 A$\cdot$GeV  favored the 'melting $\rho$'
scenario \cite{NA60}. Also the more recent data from the CERES
Collaboration (with enhanced mass resolution) \cite{CERES2} show a
preference for the 'melting $\rho$' picture.

In 2007 the PHENIX Collaboration has presented first dilepton data
from $pp$ and $Au+Au$ collisions at
Relativistic-Heavy-Ion-Collider (RHIC) energies of $\sqrt{s}$ =
200 GeV \cite{PHENIX} which show an even larger enhancement in
$Au+Au$ reactions (relative to $pp$ collisions) in the invariant
mass regime from 0.15 to 0.6 GeV than the data at
SPS energies \cite{NA60,CERES2}. The
question arises if this sizeable enhancement might also be
attributed to in-medium modifications of the $\rho$ and $\omega$
mesons as at SPS energies \cite{CBRep98,rapp5} or if new radiative
channels from the strong Quark-Gluon Plasma (sQGP) have been seen.

The answer to this question is nontrivial due to the
nonequilibrium nature of the heavy-ion  reactions and covariant
transport models have to be incorporated to disentangle the
various sources that contribute to the final dilepton spectra seen
experimentally. In this study we aim at contributing to this task
employing an up-to-date relativistic transport model (HSD) that
incorporates the relevant off-shell dynamics of the vector mesons.
The HSD transport model \cite{CBRep98,Brat97,Ehehalt} has been
used for the description of $pA$ and $AA$ collisions from SIS to
RHIC energies and lead to a fair reproduction of hadron
abundancies, rapidity distributions and transverse momentum
spectra. We recall that in the HSD approach nucleons, $\Delta$'s,
N$^*$(1440), N$^*$(1535), $\Lambda$, $\Sigma$ and $\Sigma^*$
hyperons, $\Xi$'s, $\Xi^*$'s and $\Omega$'s as well as their
antiparticles are included on the baryonic side whereas the $0^-$
and $1^-$ octet states are incorporated in the mesonic sector.
Inelastic baryon--baryon (and meson-baryon) collisions with
energies above $\sqrt s_{th}\simeq 2.6\GeV$ (and $\sqrt{s_{th}}
\simeq 2.3\GeV$) are described by the \Fritiof{} string model
\cite{FRITIOF} whereas low energy hadron--hadron collisions are
modeled in line with experimental cross sections. Low energy cross
sections such as threshold meson production in proton-neutron
($pn$) collisions - which are scarcely available from experiments
-  are fixed by proton-proton ($pp$) cross sections and isospin
factors emerging from pion-exchange diagrams. Since we address
ultrarelativistic collisions at SPS and RHIC energies such 'low
energy uncertainties' are of minor relevance here. As pre-hadronic
degrees of freedom HSD includes 'effective' quarks (antiquarks)
and diquarks (antidiquarks) which interact with cross sections in
accordance with the constituent quark model (cf. Refs.
\cite{Falter}).

Compared to our earlier studies in Refs. \cite{CBRep98,CBRW97} a
couple of extensions have been implemented such as
\begin{itemize}
\item{off-shell dynamics for vector mesons - according to
Refs. \cite{Cass_off1} - and an extended set of vector meson
spectral functions \cite{Brat08}}
\item{extension of the LUND string model to include 'modified'
spectral functions for the hadron resonances in the string decays.}
\end{itemize}
As demonstrated in Ref. \cite{Brat08} the off-shell dynamics is
particularly important for resonances with a rather long lifetime
in vacuum but strongly decreasing lifetime in the nuclear medium
(especially $\omega$ and $\phi$ mesons) but also proves vital for
the correct description of dilepton decays  of $\rho$ mesons with
masses close to the two pion decay threshold \footnote{This will
be of particular importance in comparison to the low mass dileptons
from PHENIX.}. For a detailed
description of the various hadronic channels included for dilepton
production as well as the off-shell dynamics we refer the reader
to Ref. \cite{Brat08} where we have focused on $e^+e^-$ production
in the 1 to 2 A$\cdot$GeV energy range.

\begin{figure}[t]
\phantom{a}\vspace*{5mm} \centerline{\psfig{figure=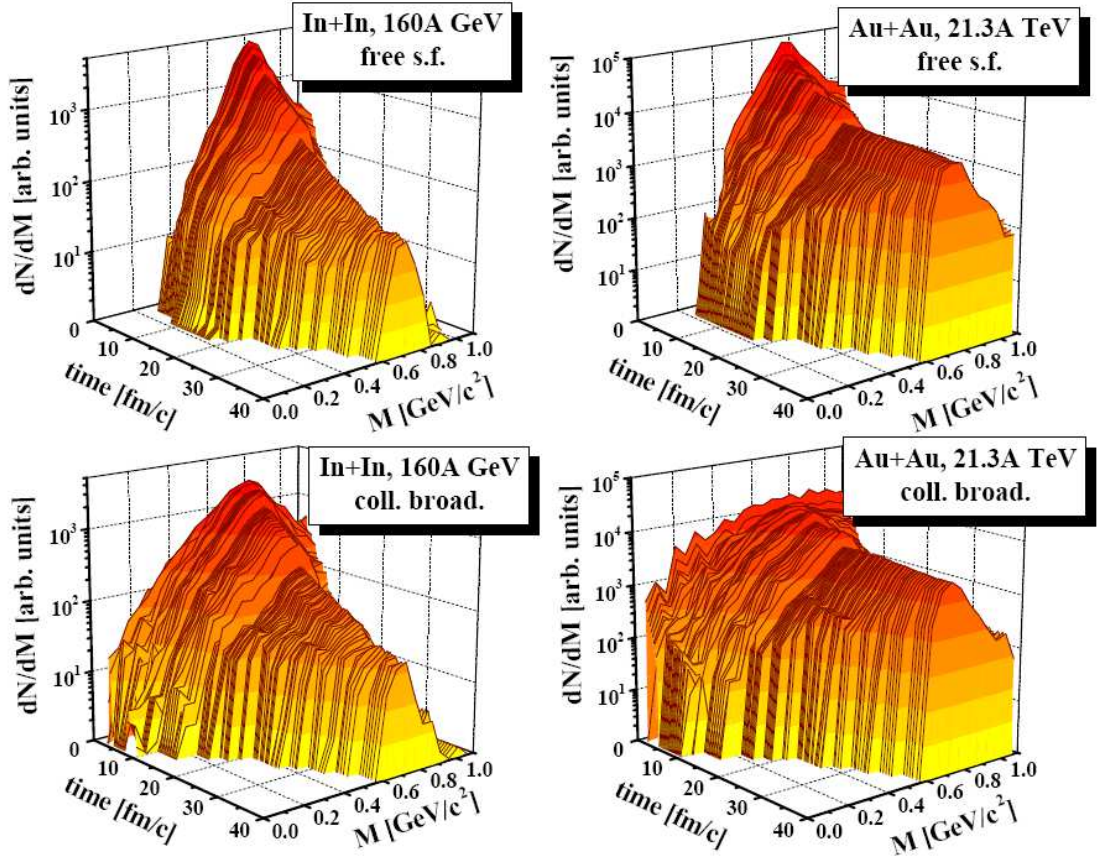,width=14.5cm}}
\caption{Time evolution of the mass distribution of $\rho$ mesons
for  central (b=0.5 fm) $In+In$ collisions at 160A~GeV (left part)
and for $Au+Au$ collisions at 21.3A~TeV (right part) for the free
spectral function (upper plots) and for the 'collisional
broadening' scenario (lower plots). } \label{Fig3D}
\end{figure}

Before we step to a comparison with experimental data we show in
Fig. \ref{Fig3D} the time evolution of the mass distribution of
$\rho$ mesons for  central (b=0.5 fm) $In+In$ collisions at
160A~GeV (left part) and for $Au+Au$ collisions at 21.3A~TeV
(right part) for the free $\rho$ spectral function (upper plots)
and the 'collisional broadening'  scenario (lower plots). In the
free case there are no mass components below $2 m_{\pi}$ whereas
in the 'collisional broadening' scenario the $\rho$ mass
distribution extends down to twice the electron mass. The $\rho$
mass distributions for times $t >$ 15 fm$/c$ essentially show the
width due to the $\rho \rightarrow \pi \pi$ decay in vacuum which
is delayed at the RHIC energy due to significantly larger Lorentz
$\gamma$ factors (with respect to the calculational frame).  In
the collisional broadening case (lower plots) we observe a
substantially larger width in the initial $\rho$ meson mass
distribution at the RHIC energy due to a larger baryon +
antibaryon density for the heavier Au+Au system and a higher
$\rho$ + meson scattering rate. The corresponding results on
dilepton spectra - within the experimental accceptance and mass
resolution - will be shown below.

We mention that also finite temperature effects lead to a sizable
broadening of the vector mesons spectral functions (also at baryon
chemical potential $\mu_B=0$). This is essentially due to vector meson
scattering with mesons which may contribute to the total width by 70-80
MeV at a temperature of $\sim 170$~MeV according to the early work by
Haglin \cite{Haglin95}.  For the present study we use a simplified
modeling of the collisional broadening by meson scattering which
discards an explicit consideration of such 'temperature effects' in the
parametrization of the vector-meson spectral functions (cf. Ref.
\cite{Brat08}).  However, the 'temperature effects' are partly
accounted here by explicit meson-meson interactions which lead to a
dynamical broadening in the vector meson mass distribution.

\begin{figure}[t]
\phantom{a}\vspace*{5mm}
{\psfig{figure=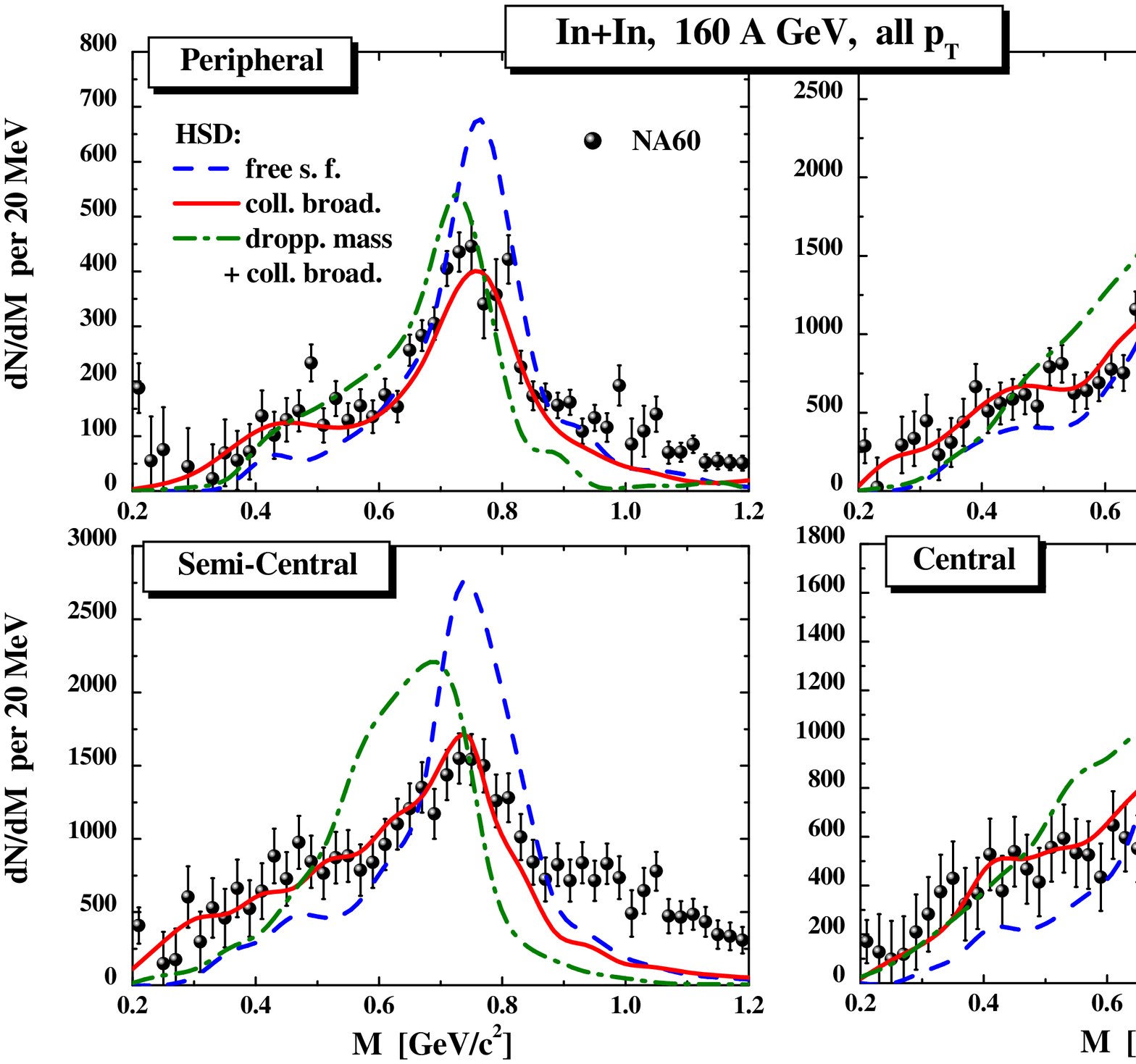,width=10.5cm}}
\caption{The HSD results for the mass differential dilepton
spectra from the direct $\rho$ meson decay  in case of $In + In$
at 158 A$\cdot$GeV for peripheral, semi-peripheral, semi-central
and central collisions in comparison to the excess mass spectrum
from NA60 \protect\cite{NA60}. The actual NA60 acceptance filter
and mass resolution have been incorporated \cite{Sanja}.  The
solid lines show the HSD results for a scenario including the
collisional bradening of the $\rho$-meson whereas the dashed
lines correspond to calculations with  'free' $\rho$ spectral
functions for reference. The dash-dotted lines represent the HSD
calculations for the 'dropping mass + collisional broadening'
model (see text).} \label{Fig1}
\end{figure}

Let's directly go over to the actual results from HSD for $In+In$
collisions at 160 A$\cdot$GeV which are displayed in Fig.
\ref{Fig1} for peripheral, semi-peripheral, semi-central and
central collisions following the centrality definition of the NA60
Collaboration \cite{NA60,NA60_05centrality,Sanja}: all measured
events have been separated into 4 centrality bins according to
their charged particle multiplicity $dN_{ch}/d\eta$ measured in
the pseudorapidity interval $3 \le \eta \le 4.2$ (cf. Ref.
\cite{NA60_05centrality} for details): bin 1 (central) - $170\le
dN_{ch}/d\eta \le 240$; bin 2 (semi-central) - $110\le
dN_{ch}/d\eta \le 170$; bin 3 (semi-peripheral) - $30\le
dN_{ch}/d\eta \le 110$; bin 4 (peripheral) - $4\le dN_{ch}/d\eta
\le 30$. By computing $dN_{ch}/d\eta$ within HSD for the same
pseudorapidity interval we obtain a direct correspondence of the
centrality bins with the impact parameter intervals: bin 1
(central) -  $b\le 3.5$ fm; bin 2 (semi-central) - $3.5\le b \le
5.5$ fm; bin 3 (semi-peripheral) - $5.5\le b \le 8.5$ fm; bin 4
(peripheral) - $b \ge 8.5$ fm which we used as centrality criteria
for our calculations. We note, furthermore, that the experimental
data in Fig. \ref{Fig1} correspond to the 'excess mass spectra',
which are extracted from the measured dilepton yields by
subtracting the 'cocktail' contribution from the $\eta, \omega,
\phi$ decays \cite{NA60}.  Such a procedure allows to 'separate'
the contribution of the $\rho$ mesons which is dominant in the
mass region $0.25 \le M \le 0.9$ GeV$c^2$ over other dilepton
sources.  Following Ref. \cite{NA60} we compare in Fig. \ref{Fig1}
the corresponding NA60 data for the excess mass spectra with the
HSD calculations for the dilepton yield from the direct $\rho$
mesons, where the normalization to the data is performed with
respect to the integral yield from 0.2 to 0.9 GeV for all scenarios
considered. Note that an
absolute normalization to the NA60 data - contrary to the systems
discussed below - is not yet available and we thus focus on a
'shape' analysis.

The dashed  lines give the dilepton mass spectra for the direct
$\rho$ meson decays when incorporating only the vacuum $\rho$-
spectral function in all hadronic reaction processes. These
reference spectra overestimate the data in the region of the
$\rho$-meson pole mass and underestimate the experimental spectra
in the region below (and above) the pole mass such that in-medium
modifications can clearly be identified. The solid lines show the
result from HSD in the 'collisional broadening' scenario where no
shift of the $\rho$ pole mass is incorporated but an increase of
the $\rho$-meson width due to hadronic collisions proportional to
the baryon density $\rho_B$ (calculated as $\rho_B^2(x) =
j^{\mu}(x) j_{\mu}(x)$ with $j^{\mu}(x)$ denoting the baryon
4-current). The dash-dotted lines represent the HSD calculations
for the 'dropping mass + collisional broadening' model where the
$\rho$  mass has been dropped with baryon density in accordance
with Eq. (10) in Ref. \cite{Brat08}. An explicit representation of
the $\rho-$ and $\omega-$ meson spectral functions employed here
is presented in Fig. 2 of Ref. \cite{Brat08}. As can be seen from
Fig. \ref{Fig1} the dilepton mass spectrum is rather well
described up to invariant masses of 0.9 GeV in the 'collisional
broadening' scenario. For higher invariant masses the data signal
additional contributions. This result is practically identical to
the calculations of van Hees and Rapp \cite{rapp3} in the
expanding fireball model - when incorporating the spectral
function from Ref.  \cite{RappNPA} - and demonstrates that the
dominant in-medium effect seen in the $\mu^+\mu^-$ spectra from
NA60 is a broadening of the $\rho$ meson. The 'dropping mass +
collisional broadening' model performs worse since it shifts too
much strength to lower invariant dilepton masses. Note, however,
that a reduced dropping of the vector-meson masses might be also
compatible with the present data sets.

In Ref. \cite{Hees08} a very detailed analysis of the NA60 data
has been performed and shown that the additional yield above 0.9
GeV partly is due to open charm decays, four-pion collisions or
'quark-antiquark' annihilation. We mention that our HSD
calculations give only a small contribution from $\pi + a_1$
collisions in this invariant mass range but a preliminary study
within the Parton-Hadron-String-Dynamics (PHSD) model
\cite{PHSDdil} suggests that - apart from open charm decays - the
extra yield seen experimentally by NA60 should be due to massive
'quark-antiquark' annihilations. Since this question is presently
open and discussed controversely
\cite{rapp12,gale1,gale2,renk12,Dusling} we concentrate on
low mass dilepton pairs in the following.

\begin{figure}[t]
\phantom{a}\vspace*{5mm} \centerline{\psfig{figure=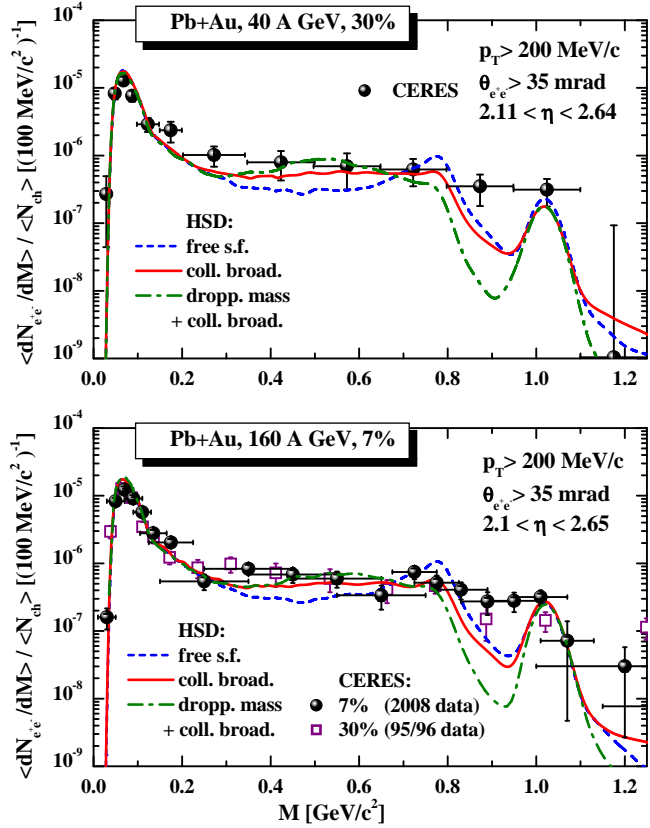,width=8.5cm}}
\caption{The HSD results  for the mass differential dilepton
spectra in 30\% central $Pb + Au$ collisions at 40 A$\cdot$GeV
(upper part) and 7\% central collisions at 158 A GeV  (lower part)
in comparison to the data from CERES \cite{CERES2}.  The dashed
lines show the results  for  vacuum spectral functions (for $\rho,
\omega, \phi$) whereas the solid lines correspond to the
'collisional broadening' scenario. The dash-dotted lines represent
the HSD calculations for the 'dropping mass + collisional
broadening' model (see text).} \label{FigCERES}
\end{figure}

The next step in our study is related to an update of the HSD
calculations in comparison to the recent data from the CERES
Collaboration \cite{CERES2} (with enhanced mass resolution).  In
Fig. \ref{FigCERES}  we present the HSD results  for the mass
differential dilepton spectra in 30\% central $Pb + Au$ collisions
at 40 A GeV (upper part) and 7\% central collisions at 158 A GeV
(lower part) in comparison to the data from CERES \cite{CERES2}.
The experimental dilepton yields in Fig. \ref{FigCERES} are
normalized to the average number of charged particles for the
corresponding centrality in the CERES acceptance: for 7\% most
central 158 A GeV events $<N_{ch}>=177$ and for 30\% central 40 A
GeV events $<N_{ch}>=216$ in the pseudo-rapidity interval $2.1\le
\eta \le 2.65$. Correspondingly, the calculated dilepton yields
have been also normalized to $<N_{ch}>$ obtained directly from the
HSD calculations which are in good agreement with the measured
numbers indicated above. The dashed lines in Fig. \ref{FigCERES}
show the results in case of vacuum spectral functions (for $\rho,
\omega, \phi$) whereas the solid lines correspond to the
'collisional broadening' scenario.  As in Fig. 2 the dash-dotted
lines represent the HSD calculations for the 'dropping mass +
collisional broadening' model. Similar to the $In+In$ case the
experimental data agree  with the calculations employing the
'collisional broadening' scenario (also in line with  Ref.
\cite{Hees08}). However, the CERES data also compare reasonably
well  with the 'dropping mass + collisional broadening' model up
to invariant masses of 0.8 GeV.

On the other hand, the HSD model underpredicts the yield between
the $\omega$ and $\phi$ peaks which might again be attributed to
possible contributions from 'quark-antiquark' annihilations etc.
We mention that in Ref. \cite{Brat97} we showed that this
invariant mass region is very sensitive to the in-medium scenario
since the simple 'dropping mass' picture provided a strong shift
of the dilepton yield to the low mass regime and a strong
reduction of the dilepton yield at $M\sim 0.9$ GeV whereas the
'collisional broadening' scenario gave a much larger contribution
at $M\sim 0.9$ GeV.  The 'dropping mass + collisional broadening'
model (dot-dashed lines) shows a similar but less pronounced
trend. Thus detailed measurements of this mass regime with high
resolution will provide interesting constraints on the in-medium
scenarios and might as well indicate possible contributions from
partonic degrees of freedom - e.g.  quark - antiquark annihilation
or gluon-Compton scattering - already at SPS energies. A more
detailed investigation within the PHSD approach will be presented
in the near future.

\begin{figure}[t]
\phantom{a}\vspace*{5mm} \centerline{\psfig{figure=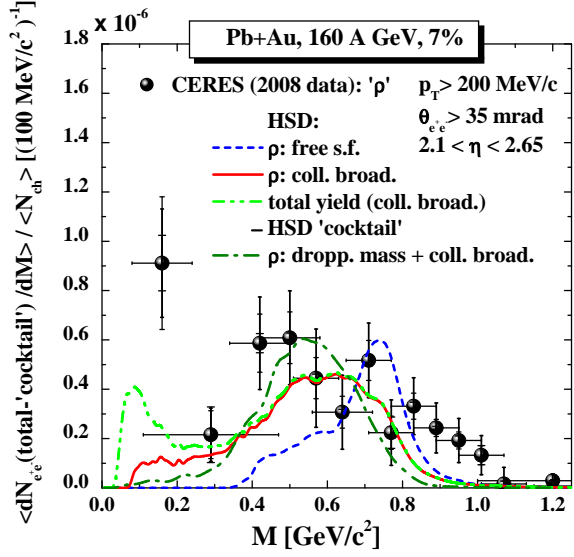,width=7.5cm}}
\caption{The comparison of the CERES data for the dilepton yield after
subtraction of the hadronic cocktail (without the $\rho$) \cite{CERES2}
with the HSD calculations:
the dashed line shows the result for the vacuum $\rho$ spectral function
whereas the solid line corresponds to the 'collisional broadening' scenario;
the dash-dot-dot line stands for the excess yield defined as
total yield after HSD 'cocktail' subtraction. The
dash-dotted line represents the HSD calculations for the
'dropping mass + collisional broadening' model.}
\label{FigCERES1}
\end{figure}

In addition to the total yield shown in  Fig. \ref{FigCERES} the
CERES Collaboration presented also the 'excess' yield (similar to
NA60) defined by the dilepton yield after subtraction of the
hadronic cocktail  \cite{CERES2}.  The hadronic 'cocktail' is
composed by the sum of $\pi^0, \eta, \omega, \eta^\prime$  Dalitz
decays and direct decays of $\omega$ and $\phi$ mesons to
dileptons. Thus, the residual dilepton yield might be attributed
to the $\rho$ meson contribution. The experimental excess yield is
shown in Fig. \ref{FigCERES1}  (solid dots with errorbars) in
comparison to the HSD calculations:  the dashed line shows the
result for the vacuum $\rho$ spectral function, the solid line
corresponds to the 'collisional broadening' scenario and the
dash-dotted line represents the HSD calculations for the 'dropping
mass + collisional broadening' model. The dash-dot-dot line stands
for the excess yield defined by the total yield (for the
'collisional broadening' scenario) after subtraction of the HSD
'cocktail'.  As seen from Fig. \ref{FigCERES1}, in  addition to
the $\rho$ meson contribution the HSD excess yield shows an
enhancement at  low invariant mass which is related to other
dilepton sources not included in the 'cocktail', in particular to
a sizeable contribution from the $\Delta$ Dalitz decay.  The HSD
excess yields show a reasonable agreement with the CERES data in
the 'collisional broadening' scenario and in the 'dropping mass +
collisional broadening' model except of the first data point,
which we relate more to uncertainties in the definition of the
'hadronic cocktail' than to new effects. More precise experimental
data at very low mass as well as improvements of the 'cocktail'
will  shed more light on this issue.

\begin{figure}[t]
\phantom{a}\vspace*{5mm} \centerline{\psfig{figure=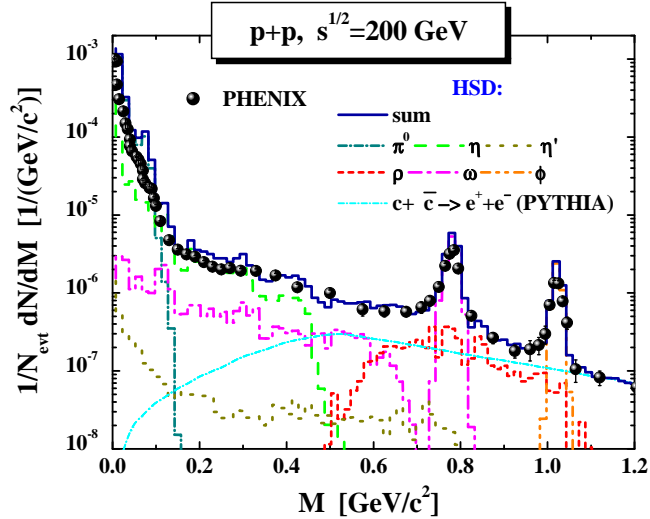,width=8.5cm}}
\caption{The HSD results  for the mass differential dilepton
spectra in case of $pp$ collisions at $\sqrt{s}$ = 200 GeV  in
comparison to the data from PHENIX \cite{PHENIXpp}. The actual PHENIX
acceptance  and mass resolution have been incorporated (see legend
for the different color coding of the individual channels). }
\label{Fig2}
\end{figure}

We step on to RHIC energies and first compare the HSD results for
the dilepton invariant mass spectrum from $pp$ collisions at
$\sqrt{s}$ = 200 GeV with the data from PHENIX \cite{PHENIXpp} in
Fig. \ref{Fig2}. The electron-positron pairs simulated in each
$pp$ event have been passed through the PHENIX acceptance and mass
resolution routines \cite{PHENIXpp}.  We note that HSD for
(inelastic) $pp$ reactions for $\sqrt{s} > $ 2.6 GeV is identical
to FRITIOF (including PYTHIA v5.5 with JETSET v7.3 for the
production and fragmentation of jets), i.e. for general hadron
production. Explicit comparisons with data for $pp$ reactions are
presented in Refs.  \cite{Brat03} as well as in Ref.
\cite{Cassing04} (Fig. 1).  Since FRITIOF is modeled (fixed) to
describe accurately elementary channels such as $pp$ or $\pi +p$
reactions also HSD performs in a similar way.

As seen from Fig. \ref{Fig2}  the HSD calculations well reproduce the
PHENIX experimental spectrum which can entirely be described by meson
Dalitz and direct decays as well as some contribution from open charm
decays (light blue thin solid line as calculated by PYTHIA). This
comparison demonstrates that the hadron production channels in HSD for
elementary $pp$ collisions are well under control also at the top RHIC
energy.

\begin{figure}[t]
\phantom{a}\vspace*{5mm} \centerline{\psfig{figure=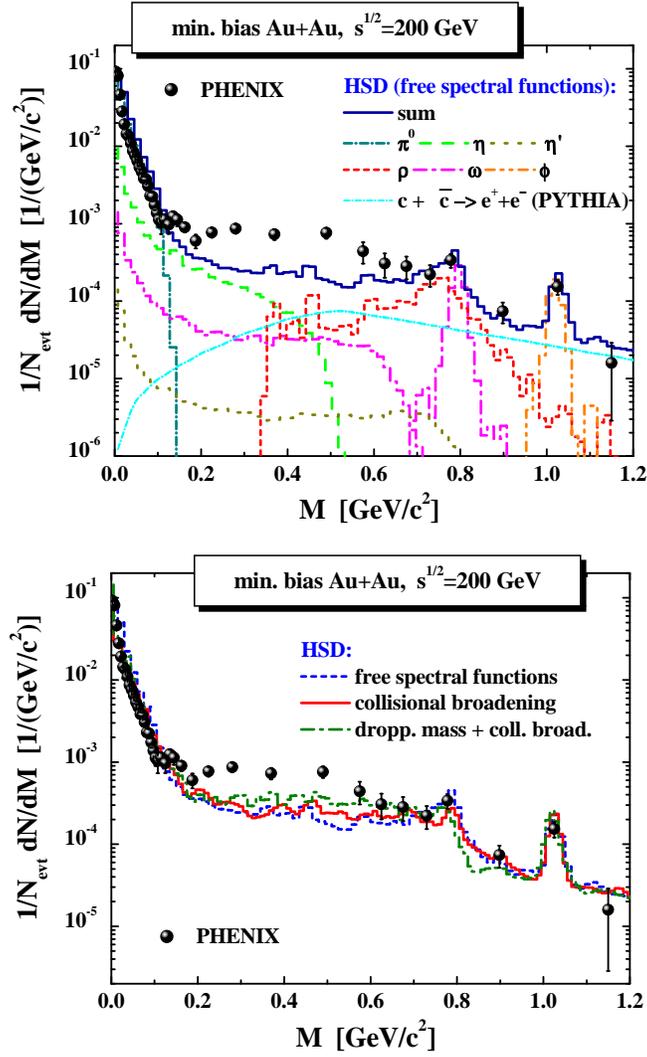,width=8.5cm}}
\caption{The HSD results  for the mass differential dilepton
spectra in case of inclusive $Au + Au$ collisions at $\sqrt{s}$ =
200 GeV  in comparison to the data from PHENIX \cite{PHENIX}. The
actual PHENIX acceptance filter and mass resolution have been
incorporated \cite{Alberica}. In the upper part the results are
shown for  vacuum spectral functions (for $\rho, \omega, \phi$)
including the channel decompositions (see legend for the different
color coding of the individual channels). The lower part shows a
comparison for the total $e^+e^-$ mass spectrum in case of the 'free'
scenario (dashed line), the 'collisional broadening' picture (solid
line) as well as the 'dropping mass + collisional broadening'
model (dash-dotted line). } \label{Fig3}
\end{figure}

We recall that HSD also provides a reasonable description of
hadron production in $Au+Au$ collisions at $\sqrt{s}$ = 200 GeV
\cite{Brat03} such that we can directly continue with the
results for $e^+e^-$ pairs which are shown in Fig. \ref{Fig3}  in
case of inclusive $Au + Au$ collisions  in comparison to the data
from PHENIX \cite{PHENIX}. Again the actual PHENIX acceptance
filter and mass resolution have been incorporated \cite{Alberica}.
In the upper part of Fig. \ref{Fig3} the results are shown for
vacuum spectral functions (for $\rho, \omega, \phi$) including the
channel decompositions (see legend for the different color coding
of the individual channels). Whereas the total yield (upper solid
line)  is quite well described in the region of the pion Dalitz
decay as well as the $\omega$ and $\phi$ mass regime we clearly
underestimate  the measured spectra in the regime from 0.2 to 0.6
GeV by an average factor of 3.

When  including the 'collisional broadening' scenario for the vector
mesons  we achieve the sum spectrum shown by the solid line in the
lower part of Fig. \ref{Fig3} which is only slightly enhanced compared
to the 'free' scenario (dashed line). Thus the question emerges if the
PHENIX data might signal dropping vector meson masses?
To answer this question we have
performed also calculations in  the 'dropping mass + collisional
broadening' model where the $\rho$ and $\omega$ masses have been
dropped with baryon density in accordance with Eq. (10) in Ref.
\cite{Brat08}. The respective HSD results are displayed in the lower
part of Fig. \ref{Fig3} by the dash-dotted  line and indeed show a
further enhancement of the dilepton yield which, however, is only small
in the mass range 0.2 GeV $< M <$ 0.4 GeV such that also this
possibility has to be excluded in comparison to the PHENIX data.

In summary we have used the off-shell version of the relativistic
HSD transport model for the calculation of dilepton spectra from
elementary as well as nucleus-nucleus collisions. Whereas the
presently available dilepton data at SIS energies of 1 to 2
A$\cdot$GeV (from the HADES Collaboration \cite{Hades1}) are well
described in the 'collisional broadening' scenario \cite{Brat08}
this also holds for low mass dimuon data from $In+In$ collisions
at 158 A$\cdot$GeV (from the NA60 Collaboration) as well as the
low mass dilepton spectra for $Pb+Au$ collisions at 40 and 158
A$\cdot$GeV (from the CERES Collaboration). The 'dropping mass +
collisional broadening' model for vector mesons seems compatible
with the CERES data at SPS energies but does not perform well for
the NA60 data.
 However, the low mass dilepton spectra from $Au+Au$ collisions at RHIC
(from the PHENIX Collaboration) are clearly underestimated in the
invariant mass range from 0.2 to 0.6 GeV in the 'collisional
broadening' scenario as well as in the 'dropping mass +
collisional broadening' model, i.e. when assuming a shift of the
vector meson mass poles with the baryon density. We mention that
our results for the low mass dileptons are very close to the
calculated spectra from van Hees and Rapp as well as Dusling and
Zahed \cite{Dussi} (cf. the comparison in Ref. \cite{AToia}).
Consequently we attribute this additional low mass enhancement
seen by PHENIX to non-hadronic sources, possibly to virtual
gluon-Compton scattering \cite{Olena}.

\section*{Acknowledgement}
The authors acknowledge inspiring discussions with
S. Damjanovic, A. Marin and A. Toia.



\begin{thebibliography}{99}
\bibitem{BrownRho}
    G.E. Brown, M. Rho, { Phys. Rev. Lett.} {\bf 66} (1991) 2720;
    Phys. Rept. {\bf 363} (2002) 85.
\bibitem{H&L92}
    T. Hatsuda, S. Lee,  { Phys. Rev.} { C}{\bf  46} (1992) R34.
\bibitem{Asakawa93}
    M. Asakawa, C.M. Ko, { Phys. Rev.} { C}{\bf 48} (1993) R526.
\bibitem{Shakin94}
    C.M. Shakin, W.-D. Sun, { Phys. Rev.}  C {\bf 49} (1994) 1185.
\bibitem{Klingl96}
    F. Klingl, W. Weise, { Nucl. Phys.}  A {\bf 606} (1996) 329;
    F. Klingl, N. Kaiser, W. Weise,
      { Nucl. Phys.} { A}{\bf  624} (1997) 527.
\bibitem{Leupold}
    S. Leupold, W. Peters, U. Mosel,
      { Nucl. Phys.} { A}{\bf 628} (1998) 311.
\bibitem{Rapp}
        R. Rapp, G. Chanfray,  J. Wambach,
       { Phys. Rev. Lett.} {\bf 76} (1996) 368.
\bibitem{Friman}
        B. Friman, H. J. Pirner,
       { Nucl. Phys.} { A}{\bf 617} (1997) 496.
\bibitem{RappNPA}
        R. Rapp, G. Chanfray, J. Wambach,
       { Nucl. Phys.} { A}{\bf 617} (1997) 472.
\bibitem{Peters}
        W. Peters, M. Post, H. Lenske, S. Leupold,  U. Mosel,
        { Nucl. Phys.} { A}{\bf 632} (1998) 109;
        M. Post, S. Leupold, U. Mosel,
       { Nucl. Phys.} { A}{\bf 689} (2001) 753.
\bibitem{CBRep98}
        W. Cassing, E. L. Bratkovskaya,
        { Phys. Rept.} {\bf 308} (1999) 65.
\bibitem{rapp5}
        R. Rapp, J. Wambach, Adv. Nucl. Phys. {\bf 25} (2000) 1.
\bibitem{CERES}
        G. Agakichiev {\it et al.}, CERES Collaboration,
              { Phys. Rev. Lett.} {\bf 75} (1995) 1272;
        Th. Ullrich {\it et al.},  { Nucl. Phys.} { A}{\bf 610} (1996) 317c;
        A. Drees, { Nucl. Phys.} { A}{\bf 610} (1996) 536c.
\bibitem{HELIOS}
        M. A. Mazzoni, HELIOS Collaboration,
              { Nucl. Phys.} { A}{\bf 566} (1994) 95c;
        M. Masera, { Nucl. Phys.} { A}{\bf 590} (1995) 93c;
        T. {\AA}kesson et al., { Z. Phys.} { C}{\bf 68} (1995) 47.
\bibitem{Li}
        G. Q. Li, C. M. Ko, G. E. Brown,
       { Phys. Rev. Lett.}  {\bf 75} (1995) 4007.
\bibitem{Li96}
        C. M. Ko, G. Q. Li, G. E. Brown,  H. Sorge,
        { Nucl. Phys.} { A}{\bf 610} (1996) 342c.
\bibitem{Cass95C}
        W. Cassing, W. Ehehalt,  C. M. Ko,
       { Phys. Lett.} { B}{\bf 363} (1995) 35.
\bibitem{Brat97}
        E. L. Bratkovskaya, W. Cassing,
      { Nucl. Phys.} { A}{\bf 619} (1997) 413.
\bibitem{Ernst}
        C. Ernst {\it et al.},
        { Phys. Rev.} { C}{\bf 58} (1998) 447.
\bibitem{CBRW97}
        W. Cassing, E. L. Bratkovskaya, R. Rapp, J. Wambach,
        { Phys. Rev.} { C}{\bf 57} (1998) 916.

\bibitem{NA60}
      R. Arnaldi {\it et al.}, NA60 Collaboration,
              Phys. Rev. Lett. {\bf 96} (2006)   162302;
      J. Seixas {\it et al.},  J. Phys. G {\bf 34} (2007) S1023;
       S. Damjanovic {\it et al.}, Nucl. Phys. {\bf A 783} (2007) 327c.
\bibitem{CERES2}
       D. Adamova {\it et al.} CERES Collaboration,
         Nucl. Phys. A {\bf 715} (2003) 262;
         Phys. Rev. Lett. {\bf 91} (2003) 042301;
       G. Agakichiev {\it et al.},
         Eur. Phys. J. C {\bf 41} (2005) 475;
       D. Adamova {\it et al.} arXiv:nucl-ex/0611022;
       A. Marin {\it et al.}; Proceedings of CPOD07,
              PoS 034 (2007), arXiv:0802.2679 [nucl-ex].
\bibitem{PHENIX}
       A. Toia {\it et al.},  PHENIX Collaboration,
       Nucl. Phys. A {\bf 774} (2006) 743;
       Eur. Phys. J {\bf 49} (2007) 243;
      S. Afanasiev {\it et al.},  arXiv:0706.3034 [nucl-ex].
\bibitem{Ehehalt}
    W. Ehehalt, W. Cassing, { Nucl. Phys.} { A }{\bf 602} (1996) 449.
\bibitem{FRITIOF}
    B. Anderson, G. Gustafson, Hong Pi,
      { Z. Phys.} { C }{\bf 57} (1993) 485.

\bibitem{Falter}
       T. Falter {\it et al.}, Phys. Lett. B {\bf 594} (2004) 61;
       Phys. Rev. C {\bf 70} (2004) 054609.
\bibitem{Cass_off1}
    W. Cassing, S. Juchem, { Nucl. Phys.} { A }{\bf 665} (2000)
    377; {\it ibid.} { A }{\bf 672} (2000) 417.


\bibitem{Brat08}
       E. L. Bratkovskaya, W. Cassing,
      Nucl. Phys. A {\bf 807} (2008) 214.

\bibitem{Haglin95}
      K. Haglin, Nucl. Phys. A {\bf 584}, 719 (1995).

\bibitem{NA60_05centrality}
      M. Floris {\it et al.}, NA60 Collaboration,
            J. Phys., Conf. Series 5 (2005) 55.

\bibitem{Sanja}
      S. Damjanovic, private communication.

\bibitem{rapp3}
       H. van Hees, R. Rapp, Phys. Rev. Lett. {\bf 97} (2006) 102301.

\bibitem{Hees08}
       H. van Hees, R. Rapp, Nucl. Phys. A {\bf 806} (2008) 339.

\bibitem{PHSDdil}
      W. Cassing, talk presented at the CERN TH Workshop on
      'Electromagnetic Radiation in Nuclear Collisions',
http://ph-dep-th.web.cern.ch/ph-dep-th/$?$site= content2/workshops/HIworkshopElmag/HIworkshopElmag.html

\bibitem{rapp12}
       R. Rapp, J. Phys. G {\bf 34} (2007) S405;
        Nucl. Phys. A {\bf 782} (2007) 275.

\bibitem{gale1}
       J. Ruppert, C. Gale, T. Renk, P. Lichard, J. I. Kapusta,
        Phys. Rev. Lett. {\bf 100} (2008) 162301.

\bibitem{gale2}
       C. Gale, S. Turbide, Nucl. Phys. A {\bf 783} (2007) 351.


\bibitem{renk12}
       J. Ruppert, T. Renk, Phys. Rev. C {\bf 71} (2005) 064903;
       Erratum-ibid. C {\bf 75} (2007) 059901;
       T. Renk, J. Ruppert,  hep-ph/0605130.

\bibitem{Dusling}
  K. Dusling, D. Teaney and I. Zahed, Phys. Rev. C {\bf 75} (2007) 024908.

\bibitem{PHENIXpp}
       A. Adare {\it et al.}, PHENIX Collaboration,
       arXiv:0802.0050 [nucl-ex].

\bibitem{Brat03}
       E. L. Bratkovskaya, W. Cassing, H. St\"ocker,
       Phys. Rev.  {\bf C 67}  (2003) 054905;
       E. L. Bratkovskaya {\it et al.}
       Phys. Rev.  {\bf C 69} (2004) 054907.

\bibitem{Cassing04}
      W. Cassing {\it et al.}, Nucl. Phys. A {\bf 735} (2004) 277.

\bibitem{Alberica} A. Toia, private communication.

\bibitem{Hades1}
       J. Pietraszko {\it et al.}, HADES Collaboration,
       Int. J. Mod. Phys. A {\bf 22} (2007) 388;
       G. Agakichiev {\it et al.},
       Phys. Rev. Lett. {\bf 98} (2007) 052302;
      T. Eberl {\it et al.}, HADES Collaboration,
              Eur. Phys. J. C {\bf 49} (2007) 261;
      G. Agakishiev {\it et al.}, Phys. Lett. B {\bf 663} (2008)
      43;
       Y.C. Pachmayer {\it et al.} arXiv:0804.3993 [nucl-ex].

\bibitem{Dussi}
      K. Dusling, I. Zahed, arXiv:0712.1982 [nucl-th].

\bibitem{AToia}
      A. Toia, Proceedings of QM2008, arXiv:0805.0153 [nucl-ex].

\bibitem{Olena}
      O. Linnyk, E. L. Bratkovskaya, W. Cassing, in preparation.



\end{thebibliography}
\end{document}